\documentclass[onecolumn,11pt]{revtex4}

\usepackage{graphicx}
\usepackage{dcolumn}
\usepackage{bm}
\usepackage{color}
\usepackage{amssymb,amsmath}

\input epsf

\begin{document}

\title{\Large{Observational Constraints of Modified Chaplygin Gas in Loop Quantum Cosmology}}

\author{\bf Shuvendu
Chakraborty$^1$\footnote{shuvendu.chakraborty@gmail.com}, Ujjal
Debnath$^2$\footnote{ujjaldebnath@yahoo.com} and Chayan
Ranjit$^1$\footnote{chayanranjit@gmail.com}} \affiliation{
$^1$Department of Mathematics, Seacom Engineering College, Howrah-
711 302, India.\\
$^2${Department of Mathematics, Bengal Engineering and Science
University, Shibpur, Howrah-711 103, India.} }

\begin{abstract}
We have considered the FRW universe in loop quantum cosmology
(LQC) model filled with the dark matter (perfect fluid with
negligible pressure) and the modified Chaplygin gas (MCG) type
dark energy. We present the Hubble parameter in terms of the
observable parameters $\Omega_{m0}$, $\Omega_{x0}$ and $H_{0}$
with the redshift $z$ and the other parameters like $A$, $B$, $C$
and $\alpha$. From Stern data set (12 points), we have obtained
the bounds of the arbitrary parameters by minimizing the
$\chi^{2}$ test. The best-fit values of the parameters are
obtained by 66\%, 90\% and 99\% confidence levels. Next due to
joint analysis with BAO and CMB observations, we have also
obtained the bounds of the parameters ($B,C$) by fixing some other
parameters $\alpha$ and $A$. From the best fit of distance modulus
$\mu(z)$ for our theoretical MCG model in LQC, we concluded that
our model is in agreement with the union2 sample data.
\end{abstract}

\maketitle

\section{\normalsize\bf{Introduction}}
The combinations of different observations astrophysical data
continuously testing the theoretical models and the bounds of the
parameters. Different observations of the SNeIa
\cite{Perlmutter,Perlmutter1,Riess,Riess1}, large scale redshift
surveys \cite{Bachall,Tedmark}, the measurements of the cosmic
microwave background (CMB) \cite{Miller,Bennet} and WMAP
\cite{Briddle,Spergel} indicate that our universe is presently
expanding with acceleration. Standard big bang Cosmology with
perfect fluid fails to accommodate the observational fact. In
Einstein's gravity, the cosmological constant $\Lambda$ (which has
the equation of state $w_{\Lambda}=-1$) is a suitable candidate
which derive the acceleration, but till now there is no proof of
the origin of $\Lambda$. Now assume that there is some unknown
matter which is responsible for this accelerating scenario which
has the property that the positive energy density and sufficient
negative pressure, know as dark energy \cite{Paddy,Sahni}. The
scalar field or quintessence \cite{Peebles} is one of the most
favored candidate of dark energy which produce sufficient negative
pressure to drive acceleration in which the potential dominates
over the kinetic term. In the present cosmic concordance
$\Lambda$CDM model the Universe is formed of $\sim$ 26\% matter
(baryonic + dark matter) and $\sim$ 74\% of a smooth vacuum energy
component. The thermal CMB component contributes only about
0.01\%, however, its angular power spectrum of temperature
anisotropies encode important information about the structure
formation process and other cosmic observables.\\

If we assume a flat universe and further assume that the only
energy densities present are those corresponding to the
non-relativistic dust-like matter and dark energy, then we need to
know $\Omega_{m}$ of the dust-like matter and $H(z)$ to a very
high accuracy in order to get a handle on $\Omega_{X}$ or $w_{X}$
of the dark energy \cite{Paddy1,Paddy2}. This can be a fairly
strong degeneracy for determining $w_{X}(z)$ from observations.
TONRY data set with the 230 data points \cite{Tonry} alongwith the
23 points from Barris et al \cite{Barris} are valid for $z>0.01$.
Another data set consists of all the 156 points in the ``gold''
sample of Riess et al \cite{Riess1}, which includes the latest
points observed by HST and this covers the redshift range $1 < z <
1.6$. In Einstein's gravity and in the flat model of the FRW
universe, one finds $\Omega_{\Lambda}+\Omega_{m}=1$, which are
currently favoured strongly by CMBR data (for recent WMAP results,
see \cite{Spergel}). In a simple analysis for the most recent
RIESS data set gives a best-fit value of $\Omega_{m}$ to be
$0.31\pm 0.04$. This matches with the value
$\Omega_{m}=0.29^{+0.05}_{-0.03}$ obtained by Riess et al
\cite{Riess}. In comparison, the best-fit $\Omega_{m}$ for flat
models was found to be $0.31\pm 0.08$ \cite{Paddy1}. The flat
concordance $\Lambda$CDM model remains an excellent fit to the
Union2 data with the best-fit constant equation of state parameter
$w=-0.997^{+0.050}_{-0.054}$(stat)$^{+0.077}_{-0.082}$(stat+sys
together) for a flat universe, or
$w=-1.038^{+0.056}_{-0.059}$(stat)$^{+0.093}_{-0.097}$(stat+sys
together) with curvature \cite{Amanullah}. Chaplygin gas is the
more effective candidate of dark energy with equation of state
$p=-B/\rho$ \cite{Kamenshchik} with $B>0$. It has been generalized
to the form $p=-B/\rho^{\alpha}$ \cite{Gorini} and thereafter
modified to the form $p=A\rho-B/\rho^{\alpha}$ \cite{Debnath}. The
MCG best fits with the 3 year WMAP and the SDSS data with the
choice of parameters $A =0.085$ and $\alpha = 1.724$ \cite{Lu}
which are improved constraints than the previous
ones $-0.35 < A < 0.025$ \cite{Jun}.\\

In recent years, loop quantum gravity (LQG) is a outstanding
effort to describe the quantum effect of our universe. Nowadays
several dark energy models are studied in the frame work of loop
quantum cosmology (LQC). Quintessence and phantom dark energy
models \cite{Wu,Chen} have been studied in the cosmological
evolution in LQC. When the Modified Chaplying Gas coupled to dark
matter in the universe is described in the frame work LQC by Jamil
et al \cite{jamil} who resolved the famous cosmic coincidence
problem in modern cosmology. In another study \cite{Fu} the
authors studied the model with an interacting phantom scalar field
with an exponential potential and deduced that the future
singularity appearing in the standard FRW cosmology can be avoided
by loop quantum effects. Here we assume the FRW universe in LQC
model filled with the dark matter and the MCG type dark energy. We
present the Hubble parameter in terms of the observable parameters
$\Omega_{m}$, $\Omega_{x}$ and $H_{0}$ with the redshift $z$. From
Stern data set (12 points), we obtain the bounds of the arbitrary
parameters by minimizing the $\chi^{2}$ test. The best-fit values
of the parameters are obtained by 66\%, 90\% and 99\% confidence
levels. Next due to joint analysis with BAO and CMB observations,
we also obtain the bounds and the best fit values of the
parameters ($B,C$) by fixing some other parameters $A$ and
$\alpha$. From the best fit of distance modulus $\mu(z)$ for our
theoretical MCG model in LQC, we concluded that our model is in
agreement with the union2 sample data.

\section{Basic Equations and Solutions for MCG in Loop Quantum Cosmology}

We consider the flat homogeneous and isotropic universe described
by FRW metric, so the modified Einstein's field equations in LQC
are given by
\begin{equation}
H^{2}=\frac{\rho}{3}\left(1-\frac{\rho}{\rho_{c}}\right)
\end{equation}
and
\begin{equation}
\dot{H}=-\frac{1}{2}(\rho+p)\left(1-\frac{2\rho}{\rho_{c}}\right)
\end{equation}
where $H$ is the Hubble parameter defined as $H=\frac{\dot{a}}{a}$
with $a$ is the scale factor. Where
$\rho_{c}=\frac{\sqrt{3}}{16\pi^{2}\gamma^{3}G^{2}\hbar}$ is
called the critical loop quantum density, $\gamma$ is the
dimensionless Barbero-Immirzi parameter. When the energy density
of the universe becomes of the same order of the critical density
$\rho_{c}$, this modification becomes dominant and the universe
begins to bounce and then oscillate forever. Thus the big bang,
big rip and other future singularities at semi classical regime
can be avoided in LQC. Let us note here it has been suggested that
$\gamma \sim 0.2375$ by the black hole thermodynamics in LQC. In
LQG, this parameter is fixed  by the requirement  of the validity
of Bekenstein-Hawking entropy for the Schwarzschild black hole.
The physical solutions are allowed only for $\rho\le\rho_{c}$. For
$\rho=\rho_{c}$, it is called bounce.  The maximum value of the
Hubble factor $H$ is reached for $\rho_{max}=\frac{\rho_{c}}{2}$
and the
maximum value of Hubble factor is $\frac{\kappa\rho_{c}}{12}$.\\

Here $\rho=\rho_{m}+\rho_{x}$ and $p=p_{x}$, where $\rho_{m}$ is
the density of matter (with vanishing pressure) and $\rho_{x}$,
$p_{x}$ are respectively the energy density and pressure
contribution of some dark energy. Now we consider the Universe is
filled with Modified Chaplygin Gas (MCG) model whose equation of
state(EOS) is given by

\begin{equation}
p_{x} = A\rho_{x}- \frac{B}{\rho_{x}^{\alpha}}, B >0 ,0\leq \alpha
\leq 1
\end{equation}
We also consider the dark matter and and the dark energy are
separately conserved and the conservation equations of dark matter
and dark energy (MCG) are given by
\begin{equation}
\dot{\rho}_{m}+3H\rho_{m}=0
\end{equation}
and
\begin{equation}
    \dot{\rho}_{x}+3H(\rho_{x}+p_{x})=0
\end{equation}

From first conservation equation (4) we have the solution of
$\rho_{m}$ as
\begin{equation}
\rho_{m}=\rho_{m0}(1+z)^{3}
\end{equation}

From the conservation equation (5) we have the solution of the
energy density as
\begin{equation}
\rho_{x}=\left[\frac{B}{A+1}+C(1+z)^{3(\alpha+1)(A+1)}\right]^{\frac{1}{\alpha+1}}
\end{equation}
where $C$ is the integrating constant, $z=\frac{1}{a}-1$ is the
cosmological redshift (choosing $a_{0}=1$) and the first constant
term can be interpreted as the contribution of dark energy. So the
above equation can be written as
\begin{eqnarray*}
\rho_{x}=\rho_{x0}\left[\frac{B}{(1+A)C+B}+\frac{(1+A)C}{(1+A)C+B}\times\right.
\end{eqnarray*}
\begin{equation}
\left.\times (1+z)^{3(\alpha+1)(A+1)}\right]^{\frac{1}{\alpha+1}}
\end{equation}
where $\rho_{x0}$ is the present value of the dark energy
density.\\

In the next section, we shall investigate some bounds of the
parameters in loop quantum cosmology by observational data
fitting. The parameters are determined by $H(z)$-$z$ (Stern), BAO
and CMB data analysis \cite{Wu1,Paul,Paul1,Paul2,Paul3}. We shall
use the $\chi^{2}$ minimization technique (statistical data
analysis) from Hubble-redshift data set to get the
constraints of the parameters of MCG model in LQC.\\

\section{\bf{Observational Data Analysis Mechanism}}

From the solution (8) of MCG and defining the dimensionless
density parameters $\Omega_{m0}=\frac{\rho_{m0}}{3 H_{0}^{2}}$ and
$\Omega_{x0}=\frac{\rho_{x0}}{3 H_{0}^{2}}$ we have the expression
for Hubble parameter $H$ in terms of redshift parameter $z$ as
follows ($8\pi G=c=1$)

\begin{eqnarray*}
H(z)=H_{0}\left[\Omega_{x0}\left(\frac{B}{(1+A)C+B}+\frac{(1+A)C}{(1+A)C+B}\times\right.\right.
\end{eqnarray*}
\begin{eqnarray*}
~~~~~~~~~~\left.\left.\times
(1+z)^{3(\alpha+1)(A+1)}\right)^{\frac{1}{1+\alpha}}
+\Omega_{m0}(1+z)^{3} \right]^{1/2}
\end{eqnarray*}
\begin{eqnarray*}
\times\left[1-\frac{3H_{0}^{2}}{\rho_{c}}\left(\Omega_{x0}\left(\frac{B}{(1+A)C+B}+\frac{(1+A)C}{(1+A)C+B}
\times\right.\right.\right.
\end{eqnarray*}
\begin{equation}
~~~~~\left.\left.\left.\times
(1+z)^{3(\alpha+1)(A+1)}\right)^{\frac{1}{1+\alpha}}
+\Omega_{m0}(1+z)^{3}\right) \right]^{1/2}
\end{equation}

From equation (9), we see that the value of $H$ depends on
$H_{0},A,B,C,\alpha,z$ so the above equation can be written as
\begin{equation}
H(z)=H_{0}E(z)
\end{equation}
where
\begin{eqnarray*}
E(z)=\left[\Omega_{x0}\left(\frac{B}{(1+A)C+B}+\frac{(1+A)C}{(1+A)C+B}\times\right.\right.
\end{eqnarray*}
\begin{eqnarray*}
~~~~~~~~~~\left.\left.\times
(1+z)^{3(\alpha+1)(A+1)}\right)^{\frac{1}{1+\alpha}}
+\Omega_{m0}(1+z)^{3} \right]^{1/2}
\end{eqnarray*}
\begin{eqnarray*}
\times\left[1-\frac{3H_{0}^{2}}{\rho_{c}}\left(\Omega_{x0}\left(\frac{B}{(1+A)C+B}+\frac{(1+A)C}{(1+A)C+B}
\times\right.\right.\right.
\end{eqnarray*}
\begin{equation}
~~~~~\left.\left.\left.\times
(1+z)^{3(\alpha+1)(A+1)}\right)^{\frac{1}{1+\alpha}}
+\Omega_{m0}(1+z)^{3}\right) \right]^{1/2}
\end{equation}

Now $E(z)$ contains four unknown parameters $A,B,C$ and $\alpha$.
Now we will fixing two parameters and by observational data set
the relation between the other two parameters will obtain and find
the bounds of the parameters.

\[
\begin{tabular}{|c|c|c|}
\hline
  ~~~~~~$z$ ~~~~& ~~~~$H(z)$ ~~~~~& ~~~~$\sigma(z)$~~~~\\
  \hline
  0 & 73 & $\pm$ 8 \\
  0.1 & 69 & $\pm$ 12 \\
  0.17 & 83 & $\pm$ 8 \\
  0.27 & 77 & $\pm$ 14 \\
  0.4 & 95 & $\pm$ 17.4\\
  0.48& 90 & $\pm$ 60 \\
  0.88 & 97 & $\pm$ 40.4 \\
  0.9 & 117 & $\pm$ 23 \\
  1.3 & 168 & $\pm$ 17.4\\
  1.43 & 177 & $\pm$ 18.2 \\
  1.53 & 140 & $\pm$ 14\\
  1.75 & 202 & $\pm$ 40.4 \\ \hline
\end{tabular}
\]
{\bf Table 1:} The Hubble parameter $H(z)$ and the standard error
$\sigma(z)$ for different values of redshift $z$.

\subsection{Analysis with Stern ($H(z)$-$z$) Data Set}

Using observed value of Hubble parameter at different redshifts
(twelve data points) listed in observed Hubble data by
\cite{Stern} we analyze the model. The Hubble parameter $H(z)$ and
the standard error $\sigma(z)$ for different values of redshift
$z$ are given in Table 1. For this purpose we first form the
$\chi^{2}$ statistics as a sum of standard normal distribution as
follows:

\begin{equation}
{\chi}_{Stern}^{2}=\sum\frac{(H(z)-H_{obs}(z))^{2}}{\sigma^{2}(z)}
\end{equation}

where $H(z)$ and $H_{obs}(z)$ are theoretical and observational
values of Hubble parameter at different redshifts respectively and
$\sigma(z)$ is the corresponding error for the particular
observation given in table 1. Here, $H_{obs}$ is a nuisance
parameter and can be safely marginalized. We consider the present
value of Hubble parameter $H_{0}$ = 72 $\pm$ 8 Kms$^{-1}$
Mpc$^{-1}$ and a fixed prior distribution. Here we shall determine
the parameters $A,B,C$ and $\alpha$ from minimizing the above
distribution ${\chi}_{Stern}^{2}$. Fixing the two parameters
$C,\alpha$, the relation between the other parameters $A,B$ can be
determined by the observational data. The probability distribution
function in terms of the parameters $A,B,C$ and $\alpha$ can be
written as

\begin{equation}
L= \int e^{-\frac{1}{2}{\chi}_{Stern}^{2}}P(H_{0})dH_{0}
\end{equation}

where $P(H_{0})$ is the prior distribution function for $H_{0}$.
We now plot the graph for different confidence levels. In early
stage the Chaplygin Gas follow the equation of state $P=A\rho$
where $A\le 1$. So, as per our theoretical model the two
parameters should satisfy the two inequalities $A\le 1$ and $B>0$.
Now our best fit analysis with Stern observational data support
the theoretical range of the parameters. The 66\% (solid, blue),
90\% (dashed, red) and 99\% (dashed, black) contours are plotted
in figures 1, 2 and 4 for $\alpha=0.5$ and $A=1,1/3,-1/3$. The
best fit values of $B$ and $C$ are tabulated in Table 2.

\[
\begin{tabular}{|c|c|c|c|}
\hline
  ~~~~~~$A$ ~~~~~& ~~~~~~~$B$ ~~~~~~~~& ~~~$C$~~~~~&~~~~~$\chi^{2}_{min}$~~~~~~\\
  \hline
  $~~1$ & 0.904 & 0.565 & 10.828 \\
  $~~\frac{1}{3}$ & 0.561 & 0.778 & 8.230 \\
  $-\frac{1}{3}$ & 0.849 & 0.599 & 7.057 \\
   \hline
\end{tabular}
\]
{\bf Table 2:} $H(z)$-$z$ (Stern): The best fit values of $B$, $C$
and the minimum values of $\chi^{2}$ for different values of $A$.

\begin{figure}
\includegraphics[height=2in]{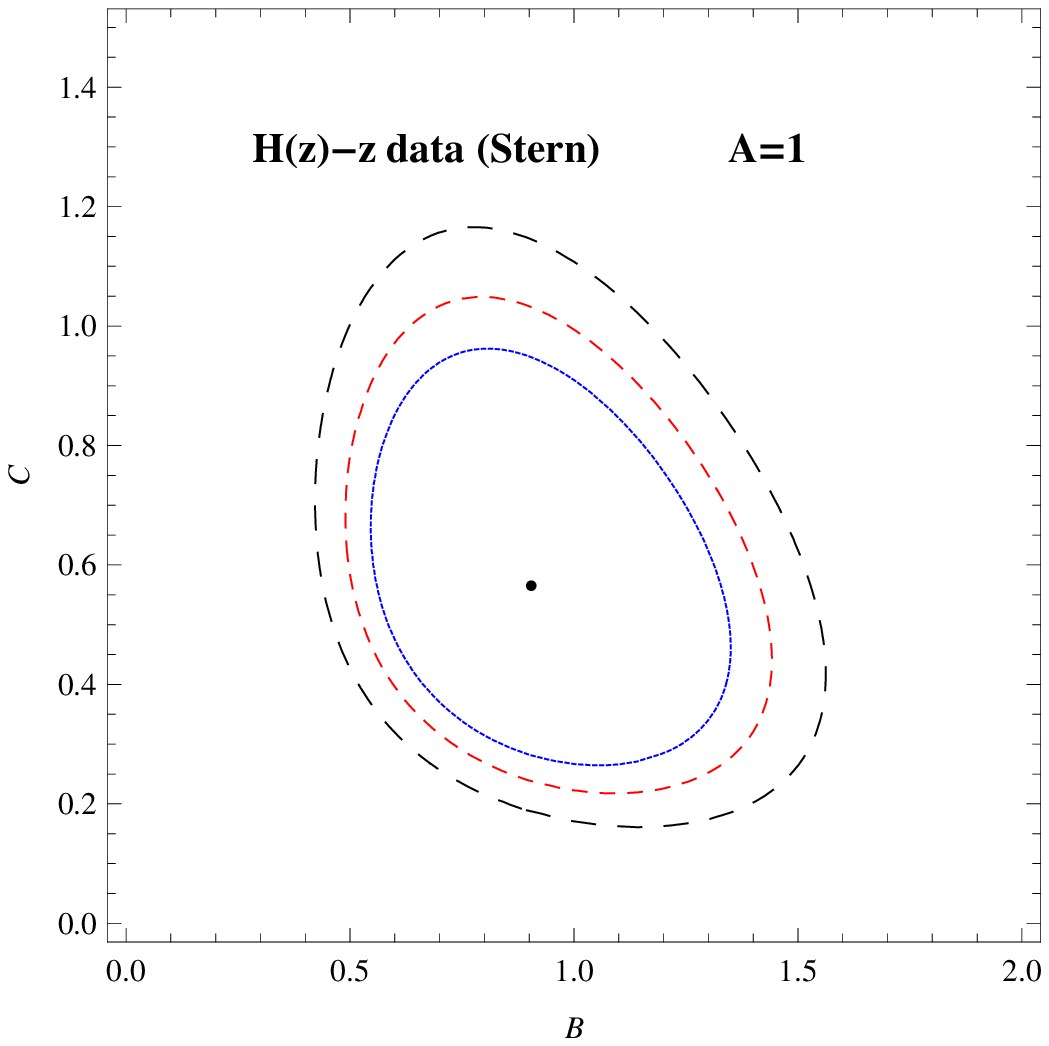}~~~~
\includegraphics[height=2in]{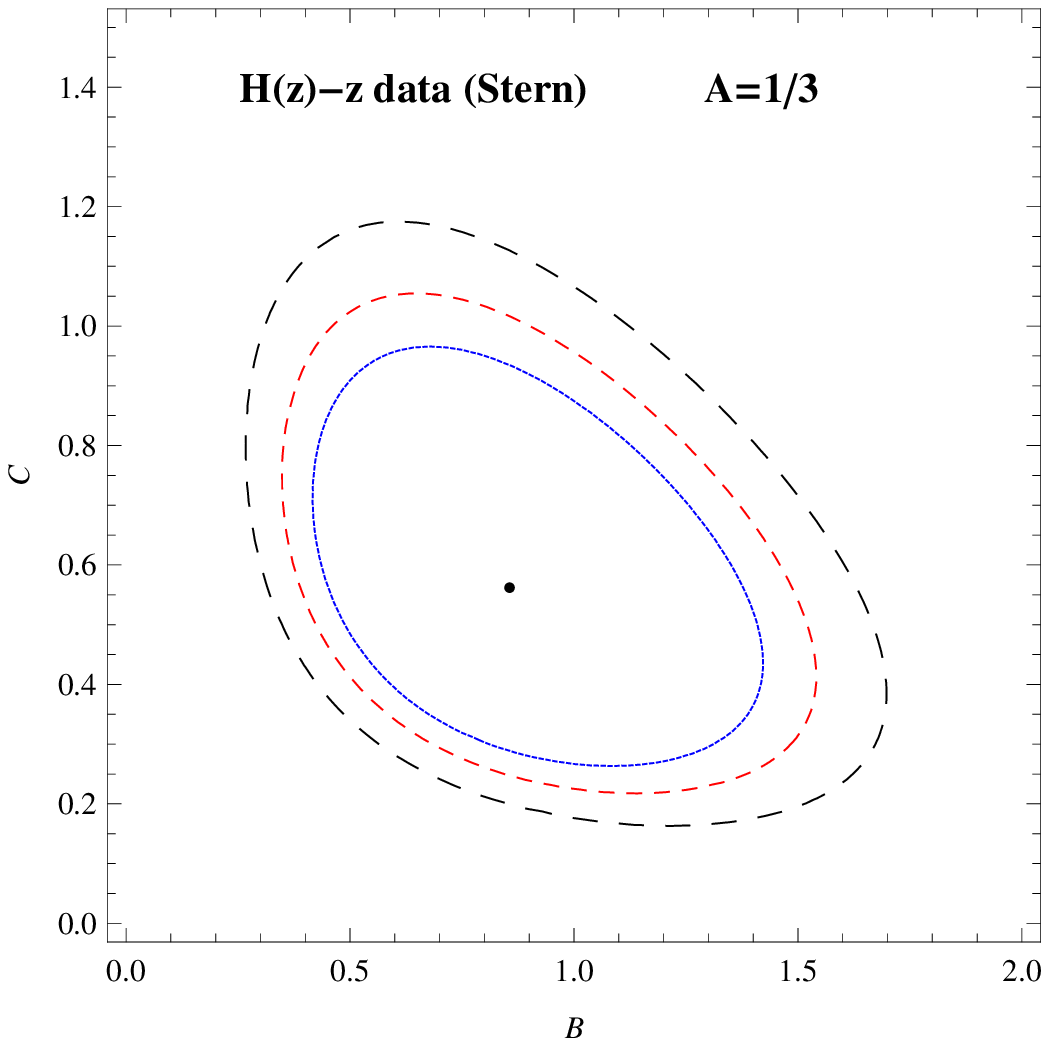}~~~~
\includegraphics[height=2in]{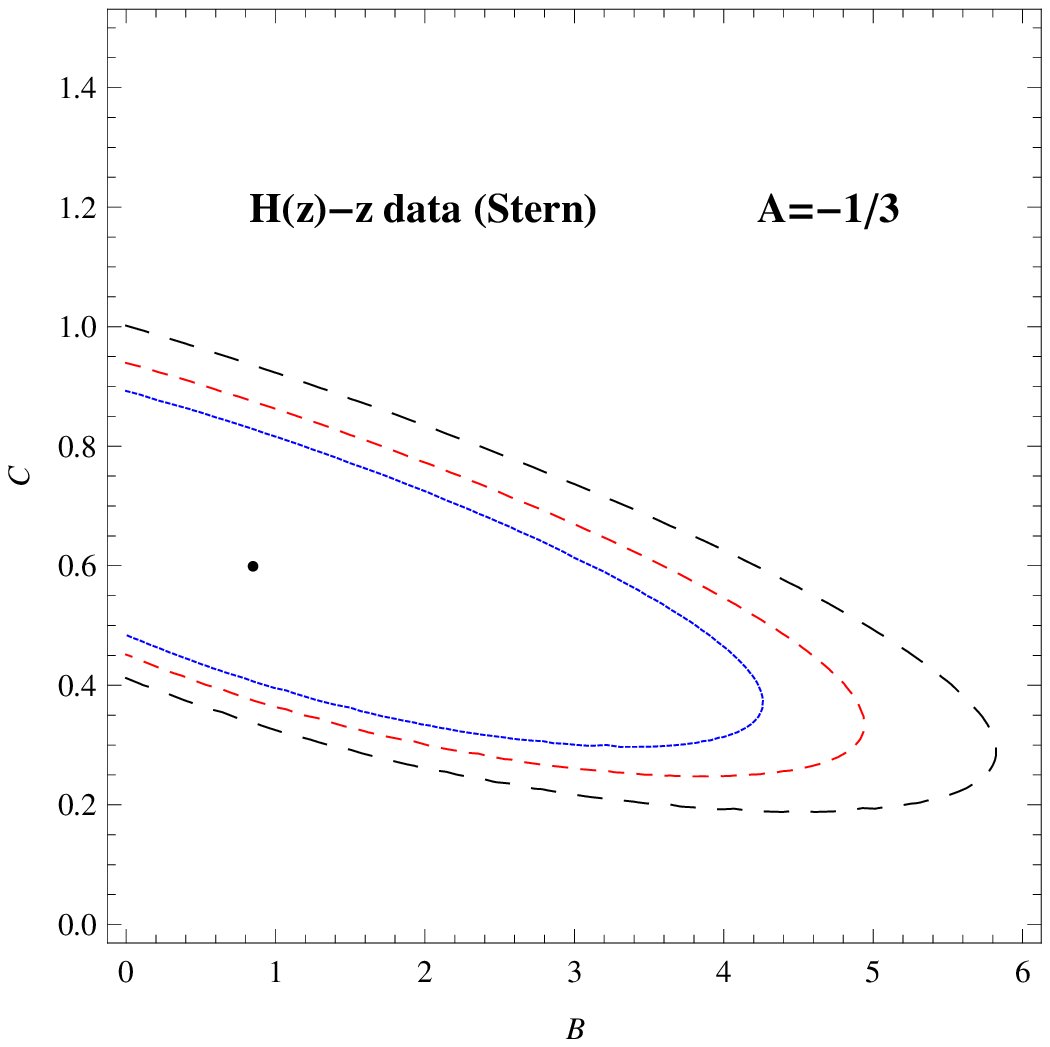}~~~~\\

Fig.1~~~~~~~~~~~~~~~~~~~~~~~~~~~~~~~~~~~~~Fig.2~~~~~~~~~~~~~~~~~~~~~~~~~~~~~~~~~Fig.3\\
\vspace{1cm}

Figs. 1 - 3 show that the variation of $C$ with $B$ for $\alpha=
0.5$, $\Omega_{0m}=0.29, \Omega_{x0}=0.72$ with $A=1,1/3,-1/3$
respectively for different confidence levels. The 66\% (solid,
blue), 90\% (dashed, red) and 99\% (dashed, black) contours are
plotted in these figures for the $H(z)$-$z$ (Stern) analysis.
\vspace{1cm}

\end{figure}

\subsection{Joint Analysis with Stern $+$ BAO Data Sets}

The method of joint analysis, the Baryon Acoustic Oscillation
(BAO) peak parameter value has been proposed by \cite{Eisenstein}
and we shall use their approach. Sloan Digital Sky Survey (SDSS)
survey  is one of the first redshift survey by which the BAO
signal has been directly detected at a scale $\sim$ 100 MPc. The
said analysis is actually the combination of angular diameter
distance and Hubble parameter at that redshift. This analysis is
independent of the measurement of $H_{0}$ and not containing any
particular dark energy. Here we examine the parameters $B$ and $C$
for Chaplygin gas model from the measurements of the BAO peak for
low redshift (with range $0<z<0.35$) using standard $\chi^{2}$
analysis. The error is corresponding to the standard deviation,
where we consider Gaussian distribution. Low-redshift distance
measurements is a lightly dependent on different cosmological
parameters, the equation of state of dark energy and have the
ability to measure the Hubble constant $H_{0}$ directly. The BAO
peak parameter may be defined by

\begin{equation}
{\cal
A}=\frac{\sqrt{\Omega_{m}}}{E(z_{1})^{1/3}}\left(\frac{1}{z_{1}}~\int_{0}^{z_{1}}
\frac{dz}{E(z)}\right)^{2/3}
\end{equation}
Here $E(z)=H(z)/H_{0}$ is the normalized Hubble parameter, the
redshift $z_{1}=0.35$ is the typical redshift of the SDSS sample
and the integration term is the dimensionless comoving distance to
the to the redshift $z_{1}$ The value of the parameter ${\cal A}$
for the flat model of the universe is given by ${\cal A}=0.469\pm
0.017$ using SDSS data \cite{Eisenstein} from luminous red
galaxies survey. Now the $\chi^{2}$ function for the BAO
measurement can be written as

\begin{equation}
\chi^{2}_{BAO}=\frac{({\cal A}-0.469)^{2}}{(0.017)^{2}}
\end{equation}

Now the total joint data analysis (Stern+BAO) for the $\chi^{2}$
function may be defined by

\begin{equation}
\chi^{2}_{total}=\chi^{2}_{Stern}+\chi^{2}_{BAO}
\end{equation}

According to our analysis the joint scheme gives the best fit
values of $B$ and $C$ in Table 3. Finally we draw the contours $A$
vs $B$ for the 66\% (solid, blue), 90\% (dashed, red) and 99\%
(dashed, black) confidence limits depicted in figures $4-6$ for $\alpha=0.5$ and $A=1,1/3,-1/3$.\\

\[
\begin{tabular}{|c|c|c|c|}
\hline
  ~~~~~~$A$ ~~~~~& ~~~~~~~$B$ ~~~~~~~~& ~~~$C$~~~~~&~~~~~$\chi^{2}_{min}$~~~~~~\\
  \hline
   $~~1$ & 0.735 & 0.610 & 827.909 \\
 $~~\frac{1}{3}$ & 0.921 & 0.735 & 768.499 \\
 $-\frac{1}{3}$ & 0.585 &0.998 & 767.440 \\
   \hline
\end{tabular}
\]
{\bf Table 3:} $H(z)$-$z$ (Stern) + BAO : The best fit values of
$B$, $C$ and the minimum values of $\chi^{2}$ for different values
of $A$.

\begin{figure}
\includegraphics[height=2in]{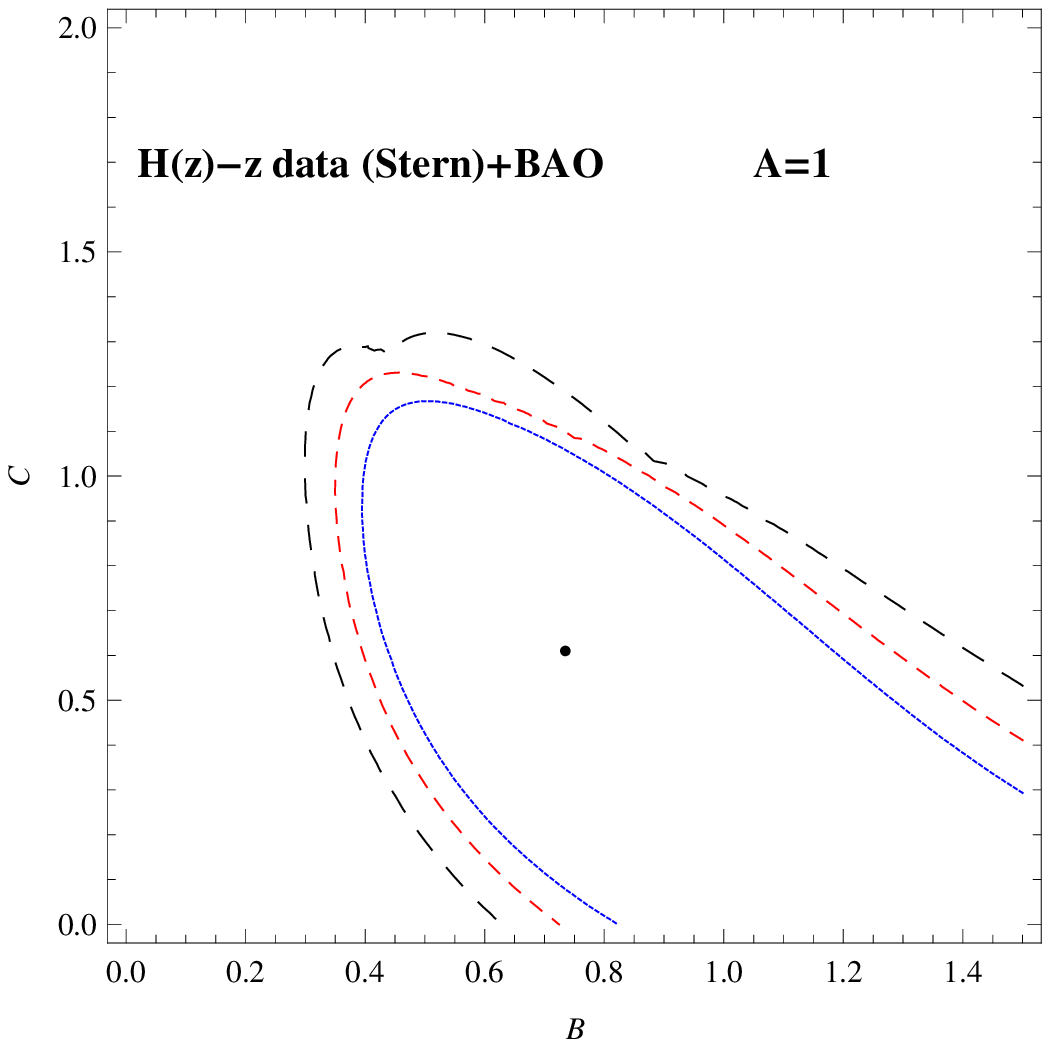}~~~~
\includegraphics[height=2in]{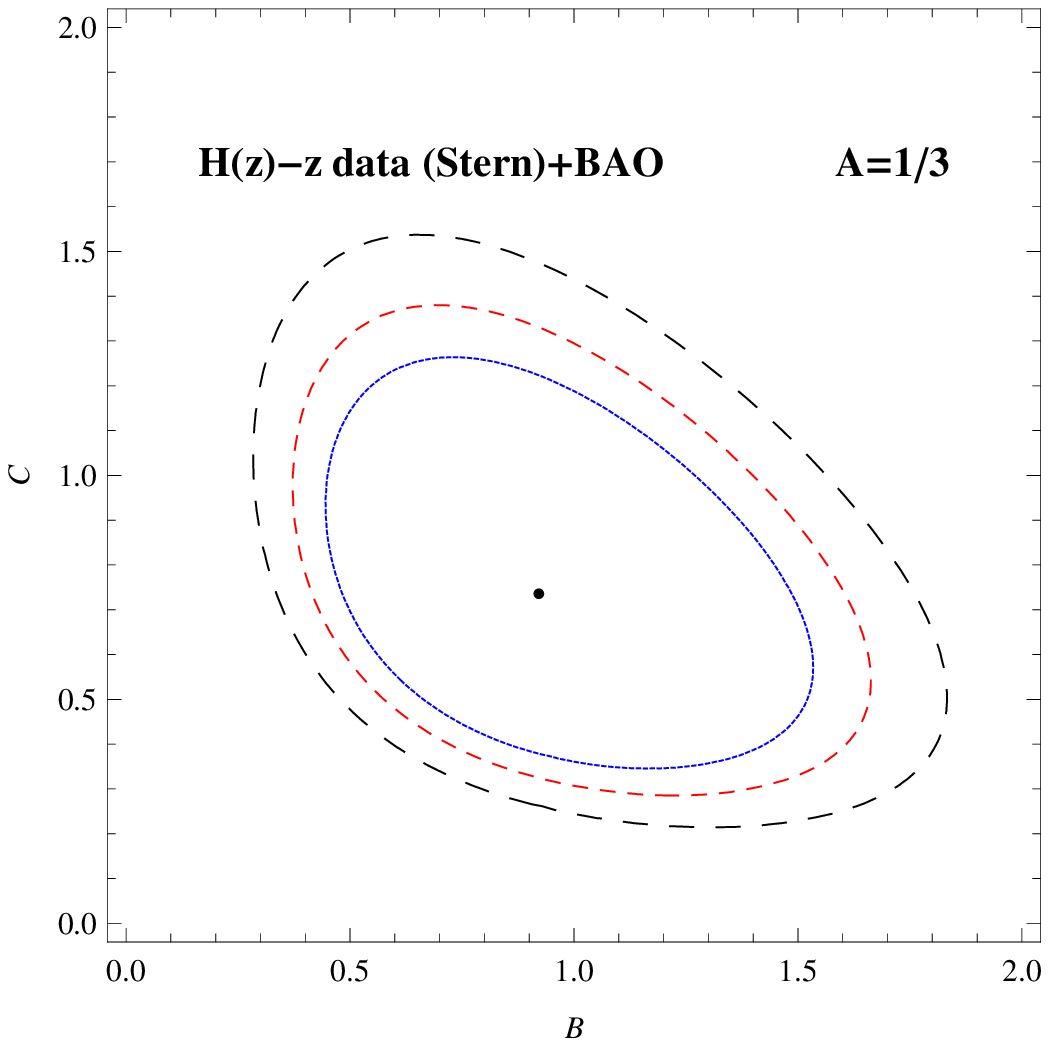}~~~~
\includegraphics[height=2in]{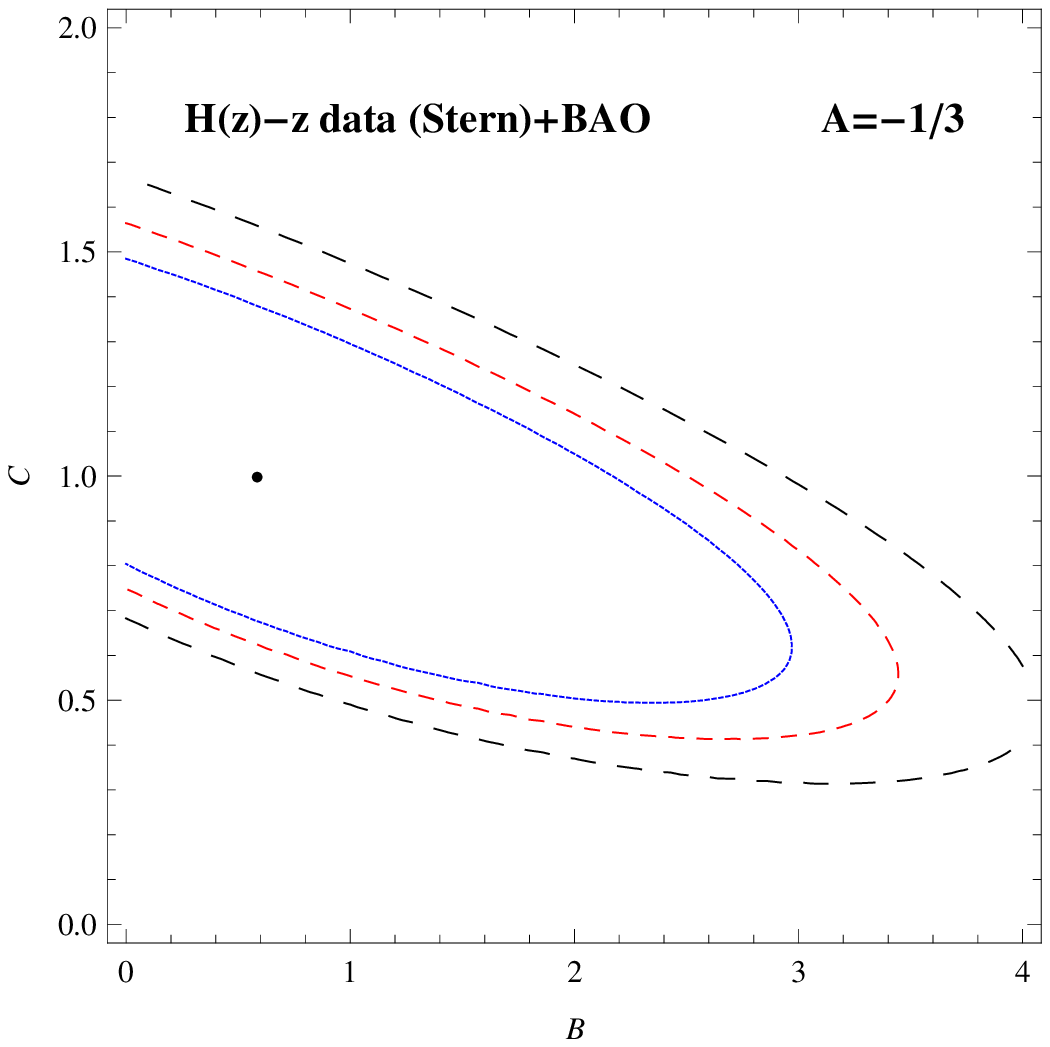}~~~~\\

Fig.4~~~~~~~~~~~~~~~~~~~~~~~~~~~~~~~~~~~~~Fig.5~~~~~~~~~~~~~~~~~~~~~~~~~~~~~~~~~Fig.6\\
\vspace{1cm}

The contours are drawn for 66\% (solid, blue), 90\% (dashed, red)
and 99\% (dashed, black) confidence levels in figs. 4 - 6 which
show the variations of $C$ against $B$ for $\alpha=0.5,
\Omega_{0m}=0.29, \Omega_{x0}=0.72$ with $A=1,1/3,-1/3$
respectively for the $H(z)$-$z$+BAO joint analysis. \vspace{1cm}

\end{figure}

\subsection{Joint Analysis with Stern $+$ BAO $+$ CMB Data Sets}

One interesting geometrical probe of dark energy can be determined
by the angular scale of the first acoustic peak through angular
scale of the sound horizon at the surface of last scattering which
is encoded in the CMB power spectrum Cosmic Microwave Background
(CMB) shift parameter is defined by
\cite{Bond,Efstathiou,Nessaeris}. It is not sensitive with respect
to perturbations but are suitable to constrain model parameter.
The CMB power spectrum first peak is the shift parameter which is
given by

\begin{equation}
{\cal R}=\sqrt{\Omega_{m}} \int_{0}^{z_{2}} \frac{dz}{E(z)}
\end{equation}

\begin{figure}
\includegraphics[height=2in]{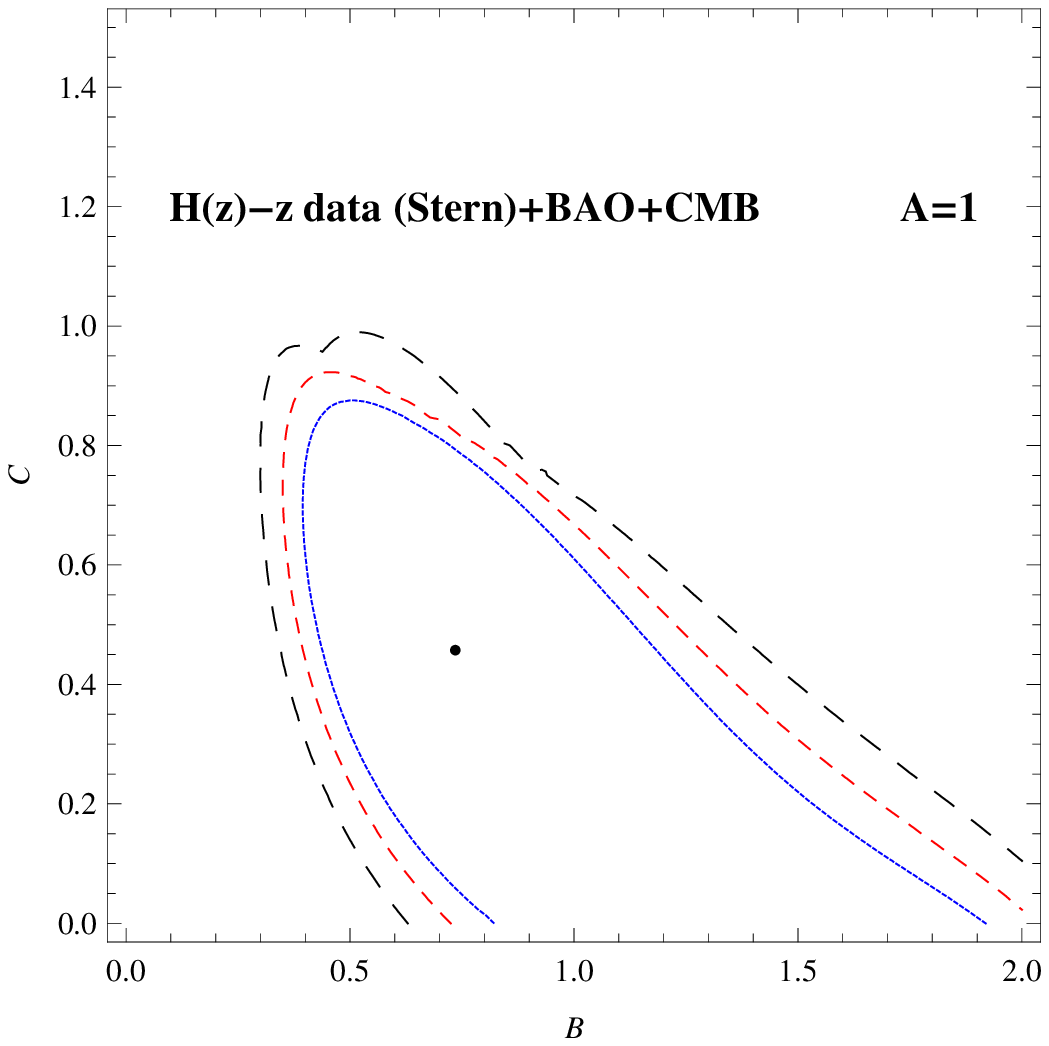}~~~~
\includegraphics[height=2in]{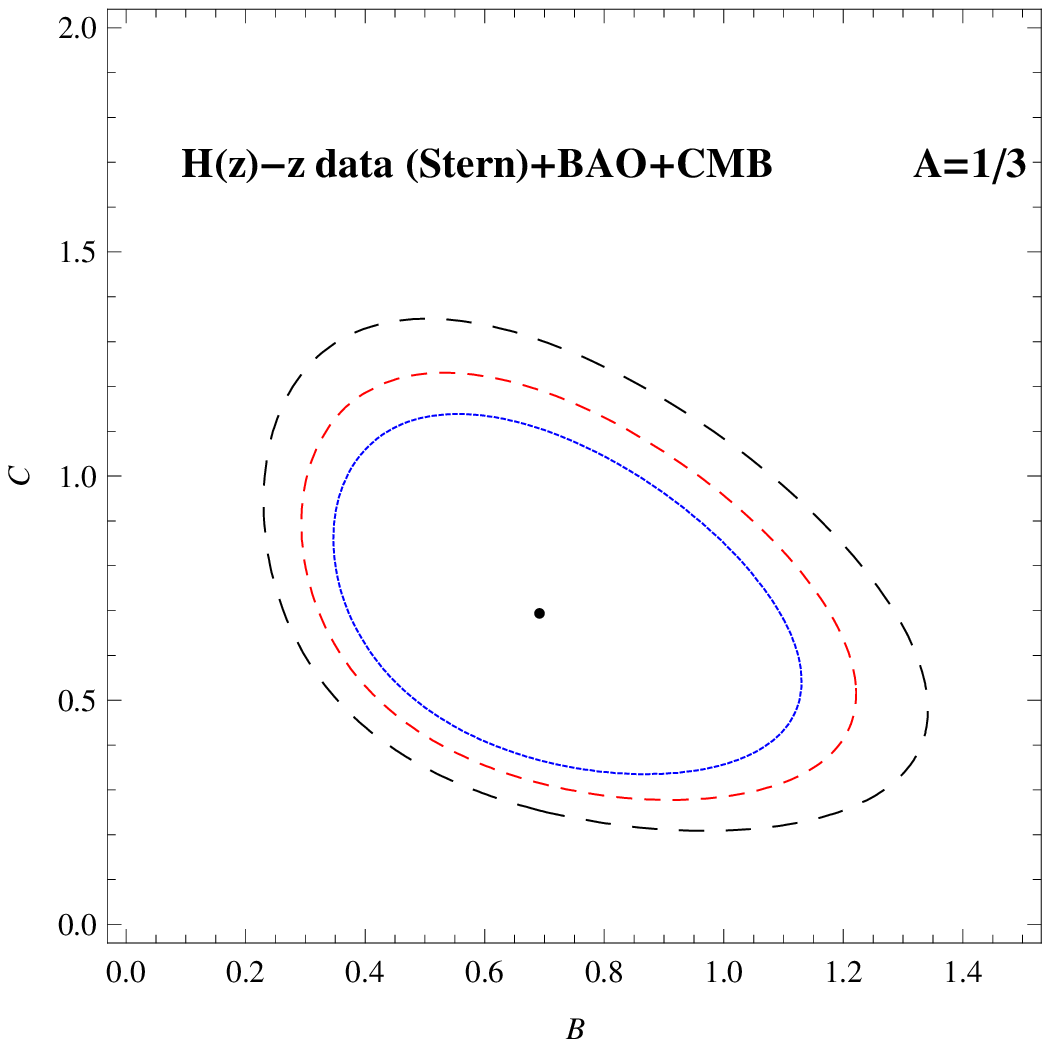}~~~~
\includegraphics[height=2in]{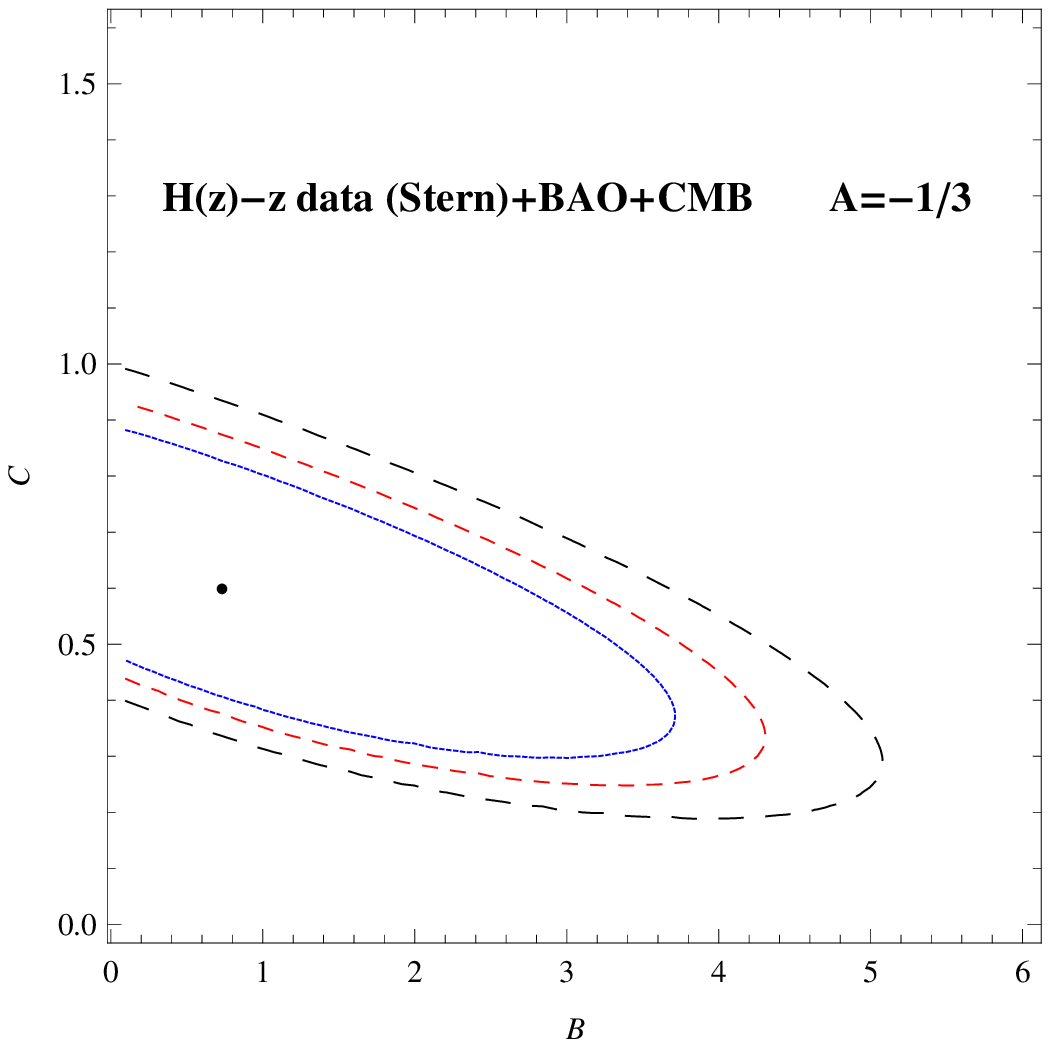}~~~~\\

Fig.7~~~~~~~~~~~~~~~~~~~~~~~~~~~~~~~~~~~~~Fig.8~~~~~~~~~~~~~~~~~~~~~~~~~~~~~~~~~Fig.9\\
\vspace{1cm}

The contours are drawn for 66\% (solid, blue), 90\% (dashed, red)
and 99\% (dashed, black) confidence levels in figs. 7 - 9 which
show the variations of $C$ against $B$ for $\alpha=0.5, C=0.01,
\Omega_{0m}=0.29, \Omega_{x0}=0.72$ with $A=1,1/3,-1/3$
respectively for the $H(z)$-$z$+BAO+CMB analysis. \vspace{1cm}

\end{figure}

where $z_{2}$ is the value of redshift at the last scattering
surface. From WMAP7 data of the work of Komatsu et al
\cite{Komatsu} the value of the parameter has obtained as ${\cal
R}=1.726\pm 0.018$ at the redshift $z=1091.3$. Now the $\chi^{2}$
function for the CMB measurement can be written as

\begin{equation}
\chi^{2}_{CMB}=\frac{({\cal R}-1.726)^{2}}{(0.018)^{2}}
\end{equation}

Now when we consider three cosmological tests together, the total
joint data analysis (Stern+BAO+CMB) for the $\chi^{2}$ function
may be defined by

\begin{equation}
\chi^{2}_{TOTAL}=\chi^{2}_{Stern}+\chi^{2}_{BAO}+\chi^{2}_{CMB}
\end{equation}
Now the best fit values of $B$ and $C$ for joint analysis of BAO
and CMB with Stern observational data support the theoretical
range of the parameters given in Table 4. The 66\% (solid, blue),
90\% (dashed, red) and 99\% (dashed, black) contours are plotted
in figures 7-9 for $\alpha=0.5$ and $A=1,1/3,-1/3$.

\[
\begin{tabular}{|c|c|c|c|}
\hline
  ~~~~~~$A$ ~~~~~& ~~~~~~~$B$ ~~~~~~~~& ~~~$C$~~~~~&~~~~~$\chi^{2}_{min}$~~~~~~\\
  \hline
  $~~1$ & 0.735 & 0.457 & 10022.6 \\
  $~~\frac{1}{3}$ & 0.692 & 0.694 & 9963.5 \\
  $-\frac{1}{3}$ & 0.731 &0.599 & 9962.11 \\
   \hline
\end{tabular}
\]
{\bf Table 4:} $H(z)$-$z$ (Stern) + BAO + CMB : The best fit
values of $B$, $C$ and the minimum values of $\chi^{2}$ for
different values of $A$.

\subsection{Redshift-Magnitude Observations from Supernovae Type Ia}

The Supernova Type Ia experiments provided the main evidence for
the existence of dark energy. Since 1995, two teams of High-$z$
Supernova Search and the Supernova Cosmology Project have
discovered several type Ia supernovas at the high redshifts
\cite{Perlmutter,Perlmutter1,Riess,Riess1}. The observations
directly measure the distance modulus of a Supernovae and its
redshift $z$ \cite{Riess2,Kowalaski}. Now, take recent
observational data, including SNe Ia which consists of 557 data
points and belongs to the Union2 sample \cite{Amanullah}.\\

From the observations, the luminosity distance $d_{L}(z)$
determines the dark energy density and is defined by

\begin{equation}
d_{L}(z)=(1+z)H_{0}\int_{0}^{z}\frac{dz'}{H(z')}
\end{equation}

and the distance modulus (distance between absolute and apparent
luminosity of a distance object) for Supernovas is given by

\begin{equation}
\mu(z)=5\log_{10}
\left[\frac{d_{L}(z)/H_{0}}{1~\text{MPc}}\right]+25
\end{equation}

The best fit of distance modulus as a function $\mu(z)$ of
redshift $z$ for our theoretical model and the Supernova Type Ia
Union2 sample are drawn in figure 10 for our best fit values of
$\alpha$, $A$, $B$ and $C$. From the curves, we see that the
theoretical MCG model in LQC is in agreement with the union2
sample data.

\begin{figure}
\includegraphics[height=2in]{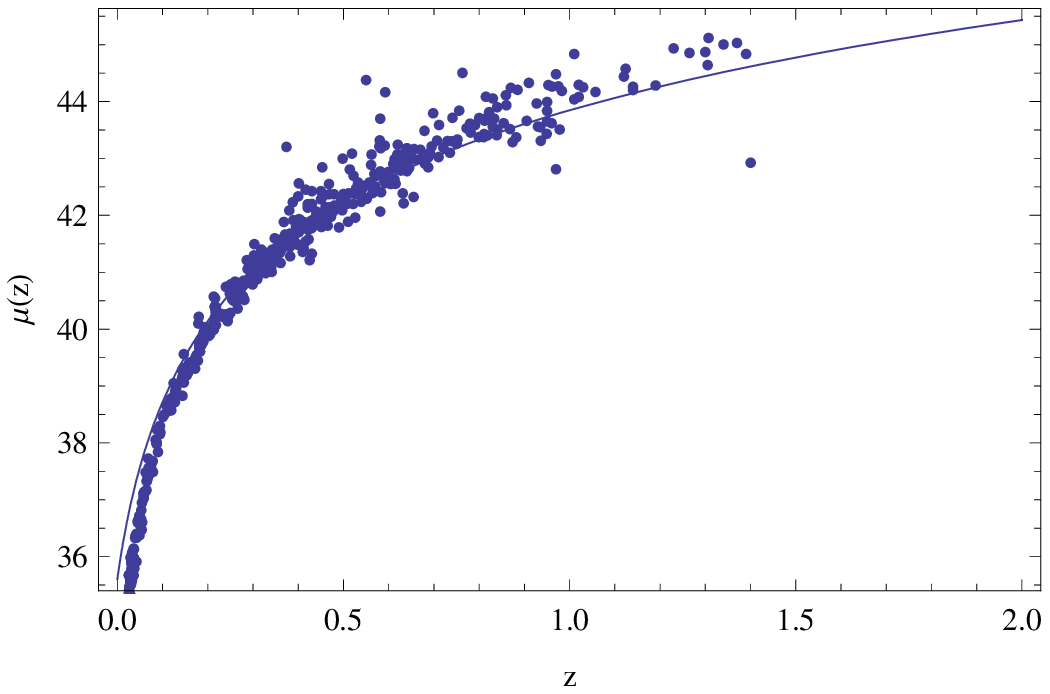}~~~~\\

Fig.10~~~~~~~~\\
\vspace{1cm}

In fig.10, $u(z)$ vs $z$ is plotted for our model (solid line) and
the Union2 sample (dotted points). \vspace{1cm}

\end{figure}

\section{\normalsize\bf{Discussions}}

Modified Chaplygin gas (MCG) is one of the candidate of unified
dark matter-dark energy model. We have considered the FRW universe
in loop quantum cosmology (LQC) model filled with the dark matter
(perfect fluid with negligible pressure) and the modified
Chaplygin gas (MCG) type dark energy. We present the Hubble
parameter in terms of the observable parameters $\Omega_{m0}$,
$\Omega_{x0}$ and $H_{0}$ with the redshift $z$ and the other
parameters like $A$, $B$, $C$ and $\alpha$. We have chosen the
observed values of $\Omega_{m0}=0.28$, $\Omega_{x0}=0.72$ and
$H_{0}$ = 72 Kms$^{-1}$ Mpc$^{-1}$. From Stern data set (12
points), we have obtained the bounds of the arbitrary parameters
by minimizing the $\chi^{2}$ test. Next due to joint analysis of
BAO and CMB observations, we have also obtained the best fit
values and the bounds of the parameters ($B,C$) by fixing some
other parameters $A=1,1/3,-1/3$ and $\alpha=0.5$. The best-fit
values and bounds of the parameters are obtained by 66\%, 90\% and
99\% confidence levels are shown in figures 1-9 for Stern,
Stern+BAO and Stern+BAO+CMB analysis. The distance modulus
$\mu(z)$ against redshift $z$ has been drawn in figure 10 for our
theoretical model of the MCG in LQC for the best fit values of the
parameters and the observed SNe Ia Union2 data sample. So our
predicted theoretical MCG model in LQC permitted the observational
data sets. The observations do in fact severely constrain the
nature of allowed composition of matter-energy by constraining the
range of the values of the parameters for a physically viable MCG
in LQC model. We have checked that when $\rho_{c}$ is large, the
best fit values of the parameters and other results of LQC model
in MCG coincide with the results of the ref. [29] in Einstein's
gravity. When $\rho_{c}$ is small, the best fit values of the
parameters and the bounds of parameters spaces in different
confidence levels in LQC distinguished from Einstein's gravity for
MCG dark energy model. Also, in particular, if we consider the
generalized Chaplygin gas ($A=0$), the best fit value of critical
Barbero-Immirzi parameter $\gamma$ is 0.2486, where we have
assumed the values of other parameters $\alpha=0.5$ and $B=0.561$
for our convenience. In summary, the conclusion of this discussion
suggests that even though the effect that quantum aspect of
gravity have on the CMB are small, cosmological observation can
put upper bounds on the magnitude of the correction coming from
quantum gravity that may be closer to the theoretical expectation
than what one would expect.\\

\section*{Acknowledgments}

The authors are thankful to IUCAA, Pune, India for warm
hospitality where part of the work was carried out.\\

\end{document}